\documentstyle[aps,preprint,epsf]{revtex}

\begin{document}

\title{A Random Matrix Model for Color Superconductivity at Zero Chemical 
Potential}

\author{Beno\^\i t Vanderheyden and A. D. Jackson\\
The Niels Bohr Institute, Blegdamsvej 17, DK-2100 Copenhagen \O, Denmark.} 

\date{\today}

\maketitle

\begin{abstract}
  We discuss random matrix models for the spontaneous breaking of both chiral
  and color symmetries at zero chemical potential and finite temperature.
  Exploring different Lorentz and gauge symmetric color structures of the
  random matrix interactions, we find that spontaneous chiral symmetry
  breaking is always thermodynamically preferred over diquark condensation.
  Stable diquark condensates appear only as $SU(2)$ rotated chiral
  condensates, which do not represent an independent thermodynamic phase.
  Our analysis is based on general symmetry arguments and hence suggests that
  no stable and independent diquark phase can form in QCD with two flavors at
  zero quark chemical potential.
\end{abstract}

\newpage

\section{Introduction}

The exploration of the phase diagram of QCD is of fundamental interest for
understanding the nature of strongly interacting matter at high temperatures
and densities and its implications for the physics of heavy-ion collisions
and neutron stars.  One central aim of studies of the phase diagram is to
determine how its global structure is shaped by underlying symmetries, their
spontaneous breaking, and their interplay.

The current picture of the QCD phase transitions is shaped by various
symmetry breaking patterns~\cite{reviews,BR}.  For QCD with two light flavors,
lattice simulations~\cite{lattice} indicate the restoration of chiral
symmetry for temperatures above $T \sim 160 $ MeV as one moves along the zero
baryon density axis.  This phase transition is of second order.  Along the
zero temperature axis, one expects chiral restoration to proceed at high
densities through a first-order transition~\cite{reviews}.  Because QCD is
asymptotically free, it is also likely that a transition to a plasma of
deconfined quarks and gluons occurs at the same high densities required for
the screening of color interactions.  The ground state of this weakly coupled
plasma may be dominated by quark-quark correlations as suggested by early and
recent works~\cite{BailinLove,NJL,instanton1,instanton2}: the attractive
quark-quark interaction may lead to the formation of quark Cooper pairs,
which spontaneously break color gauge symmetry. Nambu-Jona-Lasinio
models~\cite{NJL,NJLsuper} and instanton-based
calculations~\cite{instanton1,instanton2} suggest sizeable diquark
condensates with $\langle q q \rangle \sim 100$ MeV at quark chemical
potentials $\mu \sim 300$ MeV or higher.

Current microscopic models of color
superconductivity~\cite{BailinLove,NJL,NJLsuper,instanton1,instanton2} are
all implemented at mean-field level and, although different in their
treatment of the primary interaction among quarks, have the following
schematic phase diagram. Diquark condensates appear at the low densities
usually associated with broken chiral symmetry, but they are unstable. They
are found to be stable in the high density, chiral symmetric phase.  Between
these two limits, a state where both chiral and color symmetries are broken
appears either as a saddle point of the thermodynamical
potential~\cite{instanton1}, or as a global minimum~\cite{instanton2}. Given
the range of treatments of the QCD interactions,
it is desirable to understand the extent to which predictions of diquark
condensation are model-independent. This leads to the question of determining
how the phase diagram is shaped by global symmetries on one hand and by the
detailed dynamics of the interactions on the other.

We wish to address this question by means of random matrix methods applied to
$SU(3)$ QCD with two flavors.  As in lattice calculations, significant
complications arise from the non-Hermitean character of the Dirac operator in
models with a finite quark chemical potential. Because our main goal is to
set the stage for studying diquark condensation, we shall confine our present
exploration to Hermitean matrix models with zero chemical potential.

Before introducing our method in detail, let us briefly recall the principles
and motivations of existing random matrix models for chiral symmetry
breaking~\cite{reviewchirmm} since our formulation will follow in a similar
spirit.  Chiral random matrix models ($\chi$RMM) study the spectrum of the
Euclidean Dirac operator in a theory that respects all global symmetries of
QCD, but in which the detailed dynamics of the interaction among quarks is
replaced by averages over random matrices.  The chiral order parameter,
$<\bar \psi \psi>$, can be obtained from the density of the smallest
eigenvalues of the Dirac operator with the aid of the Banks-Casher
formula~\cite{BK}, and properties of the phase diagram can thus be
determined.  The primary advantage of such an approach lies in its
universality~\cite{universality}.  The statistical properties of the small
eigenvalues, and thus mean-field predictions regarding the chiral phase
transition, depend only on the symmetries of the Dirac operator. They do not
depend on the particular procedure used to average over the random
interactions.  In particular, chiral random matrix models reproduce the
mean-field exponents expected for QCD with two light
flavors~\cite{JacksonVerbaar}.

In this paper, we consider a random matrix model in which both chiral and
diquark condensation can take place and study the competition between these
two forms of order.  The paper is organized as follows. In
Section~\ref{s:models}, we start by constructing a class of interactions that
includes chiral and color symmetries. The dynamics of the interactions are
described by Gaussian averages over random matrices. In
Section~\ref{s:sigma}, we study the four-fermion potentials produced by
integration over all possible choices of Lorentz invariant and gauge
symmetric interactions.  Particular combinations of these choices give rise
to four-fermion interactions of the form encountered in instanton and NJL
models. In Section~\ref{s:diagram}, we introduce auxiliary variables in the
usual way, derive the thermodynamical potential, and determine the resulting
phase diagram in parameter space. As we discuss in Section~\ref{s:results},
strong dynamical constraints on the coupling parameters prevent the
exploration of certain regions of the phase diagram.  It results that the
spontaneous breaking of chiral symmetry is thermodynamically preferable to
diquark condensation in all cases.  Stable diquark condensates can appear at
most as $SU(2)$ rotated chiral condensates, which belong to a broken chiral
symmetry phase.  This case is considered in detail in section~\ref{s:axial}.
Our analysis of the phase diagram clearly distinguishes between features that
directly follow from global symmetries and those that are model-dependent. We
illustrate this point in Section~\ref{s:compare}, where we compare our phase
diagram to the predictions of NJL-model studies.

Not all the interactions to be considered here are directly related to
QCD. However, our arguments are based on simple symmetry considerations that
also apply to QCD. This suggests that no mean-field treatment of QCD with two
light flavors can find stable diquark condensates at zero chemical potential.

\section{Construction of a random matrix model}
\label{s:models}

We now turn to the construction of a random matrix model suitable for the
exploration of diquark condensation as well as chiral symmetry breaking.
Since the order parameter for color superconductivity is overall
antisymmetric in flavor, spin, and color, these quantum numbers should
explicitly be included in the model.  The extended block structure of the
resulting random matrix model is somewhat complicated.  This is an
unavoidable consequence of our desire to permit competition between two
distinct mechanisms for spontaneous symmetry breaking.  Fortunately, the
description of the QCD partition function that will emerge from our analysis
is very simple.  While we will discuss in later sections how our results
naturally follow from global symmetries, we now focus on the detailed
description of the model.

A natural starting point is to extend the $\chi$RMM~\cite{reviewchirmm} by
decomposing each chiral subblock of the single quark propagator into matrices
with a flavor, spin and color block structure. This results in a vector
theory which, as we show later, does not support diquark condensation at zero
quark chemical potential. We wish to explore further possibilities of matrix
models that build correlations in the diquark channel, and study the
resulting symmetry breaking patterns. We are forced in the process to give up
some of the symmetries of QCD. However, as we discuss later in length, our
exploration brings insight into the question of how global symmetries shape
the phase diagram. 

In the following, we only consider models for which the single quark
propagator is Hermitean. This choice ensures that the chiral and diquark
order parameters can unambiguously be related to the spectral properties of
the Dirac operator. We further impose the Dirac operator to have a $4 \times
4$ Lorentz invariant block structure in the vacuum, and a color symmetric
$N_c \times N_c$ block structure.  Once the flavor, spin, and color structure
have been specified, one is left with $N \times N$ block matrices.  It should
be stressed that the physical meaning of the matrix size $N$ here is
different from that in $\chi$RMM. Because we explicitly include color degrees
of freedom, we can no longer make analogies with instanton overlap matrices,
nor can we directly relate $N$ to the density of zero modes.  Ultimately, we
will discuss various interactions by comparing the four-point potentials
induced by the random background in the chiral and diquark channels.  Hence,
we regard the thermodynamic limit $N \to \infty$ as a way to obtain mean
field results from these potentials. The connection to zero modes is not
lost, however, as one still expects that order parameters depend only on the
dynamics of the smallest eigenvalues of the Dirac operator.

Assuming a zero vacuum angle, the finite temperature partition function has
the form
\begin{eqnarray}
Z(T) &=& \int {\cal D}H\,{\cal D}\psi_1^\dagger\,
{\cal D}\psi_1^{\phantom\dagger}
\,{\cal D}\psi_2^T\,{\cal D}\psi_2^* 
\nonumber \\
&& \times  
\exp{\left[i 
\left( \begin{array}{c}
              \psi_1^\dagger \\
              \psi_2^T
       \end{array}
\right) 
\left( \begin{array}{cc}
          {\cal H} + {\cal T} + i m & {\eta P_\Delta}\\
          -\eta^* P_\Delta^\dagger & -{\cal H}^T + {\cal T} - i m 
       \end{array}
\right)
\left( \begin{array}{c} 
       \psi_1^{\phantom\dagger} \\
       \psi_2^*
       \end{array}
\right)
\right],}
\label{partfunc}
\end{eqnarray}   
where $1$ and $2$ denote flavors and where $\psi_1^{\phantom \dagger}$,
$\psi_1^\dagger$, $\psi_2^T$, and $\psi_2^*$ are independent Grassmann
variables which are not related by complex conjugation or transposition.
${\cal H}$ is intended to mimic the interaction of a single quark with a
gluon background, and ${\cal D} H$ is the random matrix measure, which we
will write explicitly below.  We have adopted a Nambu-Gorkov representation
for the flavor structure by transposing the single quark operator for flavor
$2$.  This representation enables the random matrix interactions to build
correlations between states with the same baryon number.  We also include
external mass and diquark parameters $m$ and $\eta$. These parameters permit
the selection of a particular direction for chiral and color symmetry
breaking and are to be taken to zero in an appropriate order at the end of
the calculations. The mass $m$ is a color diagonal matrix, associated with
the chiral order parameters $\langle\psi^\dagger_1 \psi^{\phantom\dagger}_1
\rangle$ and $\langle \psi_2^T \psi^*_2 \rangle$. The complex quantity $\eta$
is to be associated with the diquark order parameters $\langle \psi^T_2
P_\Delta \psi_1^{\phantom\dagger} \rangle$ and $\langle \psi^\dagger_1
P_\Delta \psi^*_2\rangle$. We consider here diquark condensates in a $\bar 3$
state, i.e., in a scalar state which is antisymmetric in spin and in the
condensing colors.  Hence, $P_\Delta \equiv C \gamma^5 (i \lambda_2)$, where
$C$ is the charge conjugation matrix, and $i \lambda_2$ is antisymmetric in
colors $1$ and $2$.
 
The temperature $T$ enters the model in the usual
manner~\cite{JacksonVerbaar} through the chiral block matrix
\begin{eqnarray}
{\cal T}&=&\left(
                 \begin{array}{cc}
                      0         &   \pi T \\
                  \pi T       &     0 
                 \end{array}
           \right),
\label{calT}
\end{eqnarray}
where only the lowest Matsubara frequency has been retained.  In
Eq.~(\ref{partfunc}), ${\cal T}$ appears twice with a positive sign.  This
corresponds to taking opposite Matsubara frequencies for each flavor.  This
choice is justified for s-wave pairing which, in microscopic models, leads to
a homogeneous order parameter, $<\psi_2^T(x) P_\Delta
\psi^{\phantom\dagger}_1(0)>\sim \Delta$, and thus mixes Fourier components
with opposite four-momenta.

We write the Dirac operator ${\cal H}$ as an expansion into a direct product
of the 16 Dirac matrices $\Gamma^{\phantom T}_C$, times $N_c^2$ color
matrices:
\begin{eqnarray}
{\cal H}_{\lambda i \alpha k;\kappa j \beta l}  & = &
\sum_{C=1}^{16} \left( \Gamma_C^{\phantom T} \right)_{\lambda i;\kappa j} 
\sum_{a=1}^{N_c^2} \Lambda^a_{\alpha\beta} 
\left(A^{C a}_{\lambda\kappa}\right)_{k l}.
\label{calH}
\end{eqnarray}
Here, the indices ($\lambda$,$\kappa$), $(i,j)$, and $(\alpha,\beta)$
respectively denote chiral, spin, and color quantum numbers.  The indices
$(k,l)$ run from $1$ to $N$. The terms $\Lambda^a$ represent the color
matrices $\lambda^a$ when $a \le N_c^2 -1$ and the diagonal matrix
$(\delta_c)_{\alpha\beta}=\delta_{\alpha\beta}$ when $a = N_c^2$. The
normalization for color matrices is ${\rm Tr}[\lambda^a \lambda^b]=2
\delta_{ab}$ and ${\rm Tr}[\delta_c^2]=N_c$; the normalization of the Dirac
matrices is ${\rm Tr}[\Gamma_C \Gamma_{C'}]=4 \delta_{C C'}$.  In the
remainder of the paper we will explicitly keep track of $N_c$ factors, but we
will eventually be interested in the case $N_c=3$.

The random matrices $A^{Ca}_{\lambda\kappa}$ represent the fields mediating
the interaction between quarks and are thus real. Their detailed character
depends on the chiral structure of $\Gamma_C$. In a chiral basis, the Dirac
operator for vector or axial interactions has the form
\begin{eqnarray}
{\cal H}_{\rm V,A} &=&\left( 
\begin{array}{cc}
0         &        W \\
W^\dagger &        0 
\end{array}
\right),
\label{vachiral}
\end{eqnarray}
where, according to Eq.~(\ref{calH}), $W=(\Gamma_{\rm V,A})^{\phantom C}_{\rm
  RL}\, \sum_a \Lambda^a A^{Ca}_{\rm RL}$ and $W^\dagger=(\Gamma_{\rm
  V,A})^{\phantom C}_{\rm LR}\, \sum_a \Lambda^a A^{Ca}_{\rm LR}$. Since
${\cal H}$ is Hermitean, we have $A^{Ca}_{\rm LR}= (A^{Ca}_{\rm RL})^T$.
In the cases of scalar, pseudoscalar, and tensor interactions,
the Dirac operator takes the form
\begin{eqnarray}
{\cal H}_{\rm S,P,T} &=& \left(
\begin{array}{cc}
X & 0 \\
0 & Y
\end{array}
\right),
\label{sptchiral}
\end{eqnarray}
where $X=(\Gamma_{S,P,T})^{\phantom C}_{\rm RR}\, \sum_a \Lambda^a A^{Ca}$
and $Y=(\Gamma_{S,P,T})^{\phantom C}_{\rm LL}\, \sum_a \Lambda^a A^{Ca}$ are
Hermitean.  For the scalar and pseudoscalar cases and for some of the six
tensor matrices, $X$ and $Y$ are spin-diagonal and thus must be described by
real symmetric matrices $A^{Ca}$.  As a consequence of rotational invariance,
the matrices $A^{{\rm T}a}$ associated with spin off-diagonal components of
the tensor interaction must also be real symmetric.

The vector interactions mimic single-gluon exchange and respect the global
symmetries of QCD. All other interactions break at least one of them.  Axial
interactions are absent in QCD and, in a four-dimensional field theory, may
spontaneously break parity. Scalar, pseudoscalar, and tensor interactions,
break axial symmetry explicitly and may break isovector symmetry
spontaneously through the formation of $\langle \psi^\dagger_R \psi^{\phantom
  \dagger}_L \rangle$ and $\langle \psi^\dagger_L \psi_R^{\phantom \dagger}
\rangle$ condensates.  This latter condensation can actually be avoided by
retaining a finite mass parameter $m$ until the end of the calculation so
that it is always thermodynamically preferable to break symmetry in the axial
rather than vector channel. We could extend our exploration to interactions
containing the generators $\tau_i$ of $SU(2)$ flavor and study the ensuing
effective four-fermion potentials. Here, however, we will not adopt such an
approach.  It will be seen from the discussion below that the addition of
further flavor symmetry will not change our conclusions.

Having specified the random matrices $A^{Ca}_{\lambda\kappa}$, we now define
the measure ${\cal D} H$
\begin{eqnarray}
{\cal D} H &=& \left\{\prod_{C a}\prod_{\lambda \kappa} 
{\cal D}A^{Ca}_{\lambda\kappa}\right\} \,\,
\exp\left[ - N \sum_{C a} \sum_{\lambda\kappa}\,\beta_C\Sigma_{Ca}^2
\,\,{\rm Tr} [A^{Ca}_{\lambda\kappa} (A^{Ca}_{\lambda\kappa})^T]
    \right],
\label{measure}
\end{eqnarray}
where ${\cal D}A^{Ca}_{\lambda \kappa}$ are Haar measures. The index
$\beta_C$ is $\beta_C=1$ for real matrices ($C=V,A$) and $\beta_C=1/2$ for
real symmetric matrices ($C=S,P,T$).  In order to respect color symmetry, the
members of each $N_c^2-1$ color multiplet share the same variance.  In other
words, $\Sigma_{Ca}=\Sigma_{C|O}$ for $1 \le a \le N_c^2-1$, and
$\Sigma_{Ca}=\Sigma_{C|S}$ for the singlet component ($a=N_c^2$).  Further,
once the color channel $a=S,O$ has been chosen, we respect Lorentz invariance
in the vacuum by requiring a common variance $\Sigma_{T|S,O}$ for each of the
six components of the tensor interaction.  The four components of the vector
and axial interactions are characterized by the variances $\Sigma_{V|S,O}$
and $\Sigma_{A|S,O}$.

\section{Towards a non-linear sigma model}
\label{s:sigma}

We can solve the model of Eq.~(\ref{partfunc}) exactly by standard methods.
The first step is to perform the Gaussian integration over the matrix
elements. This leads to a four-fermion interaction ${\cal Y}$ and a partition
function of the form
\begin{eqnarray}
Z(T) & = & \int {\cal D}\psi_1^\dagger\,{\cal D}\psi_1\,
{\cal D}\psi^T_2\,{\cal D}\psi^*_2 \,\,
\exp{\left[ {\cal Y} + i 
\left( \begin{array}{c}
              \psi_1^\dagger \\
              \psi_2^T
       \end{array}
\right) 
\left( \begin{array}{cc}
          {\cal T}+i m & \eta P_\Delta \\
         - \eta^* P_\Delta^\dagger         & {\cal T}- i m 
       \end{array}
\right)
\left( \begin{array}{c} 
       \psi_1^{\phantom \dagger} \\
       \psi_2^*
       \end{array}
\right)
\right]}.
\label{partfuncY}
\end{eqnarray}
By introducing auxiliary variables, one can then derive a non-linear sigma
model, the saddle point of which provides the exact solution to the original
model in the limit $N\to \infty$.  In this section, we concentrate on the
fate of the diquark and the chiral condensates and thus consider only the
relevant auxiliary variables.  While it is largely a technical matter, it is
important to note that the four-point interaction has a slightly different
structure depending on whether the interaction is either vector or axial or
is either scalar, pseudoscalar, or tensor.  We now discuss these two cases
separately.

\subsection{Vector and axial interactions $(C=V,A)$}

Integrating out the real matrices $A^{Ca}_{\rm RL}$ produces a four-point
interaction of the form\footnote{ Note that the product ${\rm Tr}[A^{Ca}_{RL}
  A^{Ca}_{LR}]$ appears twice in the measure; hence the factor $8$.}
\begin{eqnarray}
{\cal Y}_{C} &=& -\sum_{\mu a}{1\over {8 N \Sigma_{Ca}^2}} 
\sum_{k,l=1}^N (J^{\mu a}_{kl})^2,
\label{Yva}
\end{eqnarray}
where the quark current $J^{\mu a}$ is 
\begin{eqnarray}
J^{\mu a}_{kl}=\psi^\dagger_{1Rk}\,\Gamma^{\mu}_{RL} \Lambda^{a}
\,\psi^{\phantom\dagger}_{1Ll}&+&
\psi^\dagger_{1Ll}\,\Gamma^{\mu}_{LR} \Lambda^{a}\,
\psi^{\phantom\dagger}_{1Rk}
\nonumber \\
& - &
\psi^T_{2Rk}\,(\Gamma^{\mu}_{LR})^T (\Lambda^{a})^T\,\psi^*_{2Ll}-
\psi^T_{2Ll}\,(\Gamma^{\mu}_{RL})^T (\Lambda^{a})^T\,\psi^*_{2Rk}
\label{vacurrent}
\end{eqnarray}
and where $\Gamma^\mu=\gamma^\mu$ and $\Gamma^\mu= i \gamma^\mu \gamma^5$ for
vector and axial interactions, respectively.  When taking the square of
$J^{\mu a}_{kl}$ in Eq.~(\ref{Yva}), we need retain only those cross-terms
relevant to chiral and diquark condensates. This becomes clear after suitable
Fierz transforms of the Lorentz and color operators.  Consider, for example,
the four-point interaction resulting from the $N_c^2-1$ color multiplet,
$\Lambda^a=\lambda^a$. Denoting the Fierz coefficient in the chiral channel
by $f^O_\chi$, the cross-term between the first and second terms in
Eq.~(\ref{vacurrent}) gives
\begin{eqnarray}
\sum_{\mu=1}^4 \sum_{a=1}^{N_c^2-1} 
(\Gamma^\mu_{RL})^{\phantom a}_{ij}\,\lambda^a_{\alpha\beta} \times 
(\Gamma^\mu_{LR})^{\phantom a}_{i'j'}\,\lambda^a_{\alpha'\beta'}
&=& f^{O}_{\chi} \,\, \delta_{ij'} \delta_{\alpha\beta'} \times
\delta_{i'j}\delta_{\alpha'\beta} + \cdots ,
\label{FierzRL}
\end{eqnarray}
and leads to a four-point interaction of the form 
\begin{eqnarray}
{\cal Y}\sim +\,\,{f^O_\chi \over {4 N \Sigma_{C|O}^2}} \,\,\,
\sum_k \psi^\dagger_{1Rk} \psi^{\phantom \dagger}_{1Rk}\times
\sum_l \psi^\dagger_{1Ll} \psi^{\phantom \dagger}_{1Ll} \ ,
\label{Yfchi}
\end{eqnarray}
which is relevant for the formation of a chiral condensate $<\psi^\dagger
\psi^{\phantom \dagger}>$.  Similarly, the cross-term between the first and
fourth terms in Eq.~(\ref{Yva}) can be projected onto a product of diquark
bilinears by means of the transformation
\begin{eqnarray}
- \sum_{\mu=1}^4 \sum_{a=1}^{N_c^2-1} 
(\Gamma^\mu_{RL})^{\phantom T}_{ij}\, \lambda^a_{\alpha\beta} \times
(\Gamma^\mu_{RL})^T_{i'j'}\, (\lambda^a)^T_{\alpha'\beta'}
&=& f^{O}_{\Delta}\,\, \varepsilon_{ij'}\,(i \lambda_2)_{\alpha\beta'} \times
\varepsilon_{i'j}\,(i \lambda_2)_{\alpha'\beta}+ \cdots,
\label{FierzRLT}
\end{eqnarray}
where $f^O_\Delta$ is now the Fierz coefficient in the diquark channel, while
$\varepsilon=(C \gamma^5)^{\phantom T}_{\rm RR}= (C \gamma^5)^{\phantom
  T}_{\rm LL}$ and $(i \lambda_2)_{\alpha\beta}$ are respectively spin and
color antisymmetric tensors. This transformation leads to the four-fermion
interaction
\begin{eqnarray}
{\cal Y} \sim + \,\, {f^O_\Delta \over {4 N \Sigma_{C|O}^2 }} \,\,\,\sum_k
\psi^\dagger_{1Rk} \, (i \,\varepsilon\, \lambda_2) \, \psi^{*}_{2Rk} \times
\sum_l
\psi^T_{2Ll} \, (i \,\varepsilon \, \lambda_2) \, \psi^{\phantom T}_{1Ll},
\label{YfDelta}
\end{eqnarray}
which is relevant for the formation of a diquark condensate $<\psi^T_2
P_\Delta \psi_1^{\phantom T}>$.

Transposing the Dirac operators in Eqs.~(\ref{FierzRL}) and (\ref{FierzRLT})
provides us with the relations needed to transform the cross-terms of the
third and fourth terms and of the second and third terms in
Eq.~(\ref{vacurrent}). Taking into account the contributions from the
$N_c^2-1$ color multiplet and the color singlet leads then to the total
four-point interaction
\begin{eqnarray}
{\cal Y}_{C}&=& + {1 \over {4 N}} \sum_{a=O,S} 
{1\over \Sigma_{C|a}^2}\sum_{k,l}
\left\{ f^a_{\chi}
\left(
\psi^\dagger_{1Rk}\psi^{\phantom\dagger}_{1Rk}\times
\psi^\dagger_{1Ll}\psi^{\phantom \dagger}_{1Ll}+
\psi^T_{2Rk}\psi^*_{2Rk}\times\psi^T_{2Ll}\psi^*_{2Ll} 
\right) 
\right.
\nonumber \\
& & + 
\left. f^a_{\Delta}
\left(
\psi^\dagger_{1Rk}\, (i \varepsilon \lambda_2)\, \psi^*_{2Rk}\times
\psi^T_{2Ll}\, (i \varepsilon \lambda_2) \, \psi_{1Ll}^{\phantom T} +
\psi^\dagger_{1Ll} \, (i \varepsilon \lambda_2) \, \psi^*_{2Ll}\times
\psi^T_{2Rk}\, (i \varepsilon \lambda_2) \, \psi^{\phantom T}_{1Rk}\,\,
\right)
\right\}.
\label{Yvatotal}
\end{eqnarray}

\subsection{Scalar, pseudoscalar, and tensor interactions ($C=S,P,T$)}
 
The random matrices representing the interaction in the scalar, pseudoscalar
and tensor channels are real symmetric. Their integration produces the
four-fermion interaction
\begin{eqnarray}
{\cal Y}_{\rm C} &=& - \sum_{a \mu} {1 \over {16 N \Sigma^2_{C|a}}}
\sum_{kl} \left( J^{a \mu}_{kl}+ J^{a\mu}_{lk} \right)^2,
\label{Yspt}
\end{eqnarray}
where $a$ again denotes the color channel ($a=O,S$) and, in the case of a
tensor interaction, $\mu=1, \ldots, 6$ labels the six antisymmetric Dirac 
matrices. The quark current here is 
\begin{eqnarray}
J^{a \mu}_{kl}&=& \psi^\dagger_{1Rk}\,\, (\Gamma_{C}^{\mu})_{\rm RR}
\Lambda^a \,\, \psi^{\phantom\dagger}_{1Rl}
- \psi^T_{2 R l}\,\, (\Gamma_{C}^{\mu})^T_{\rm RR} (\Lambda^a)^T\,\,
\psi^*_{2Rk} + \left\{ {\rm R} \leftrightarrow {\rm L} \right\}.
\label{sptcurrent}
\end{eqnarray}
Once again, only some of the cross-terms in Eq.~(\ref{Yspt}) contribute
to the chiral and diquark channels. After the necessary Fierz transforms,
these terms lead to the four-fermion interaction
\begin{eqnarray}
{\cal Y}_{\rm C}&=& \sum_{a} {1\over {8 N \Sigma^2_{C|a}}}\sum_{k,l}
\left\{
f^a_\chi \left( (\psi^\dagger_{1Rk} \psi^{\phantom\dagger}_{1Rk})^2+
(\psi^T_{2Rk} \psi^*_{2Rk})^2 \right)  \right. \nonumber \\
&& \left.
+ 2 \,\,
f^a_\Delta \,\,\psi^\dagger_{1Rk} \, (i \varepsilon \lambda_2) \, \psi^*_{2Rk}
\times \psi^T_{2Rl} \, (i \varepsilon \lambda_2) \, \psi_{1Rl}^{\phantom T} 
+\left\{ {\rm R} \leftrightarrow {\rm L} \right\}
\right\}.
\label{Yspttotal}
\end{eqnarray}

\section{Phase diagram in parameter space}
\label{s:diagram}

We are now in a position to derive a non-linear sigma model from the total
four-point interaction, ${\cal Y}=\sum_{C} {\cal Y}_C$ with ${\cal Y}_C$
given by Eqs.~(\ref{Yvatotal}) and (\ref{Yspttotal}). We first write ${\cal
  Y}$ as the sum of squares of fermion bilinears and then transform the
quartic terms into fermion bilinears by means of the Hubbard-Stratonovich
formula
\begin{eqnarray}
\exp(-A Q^2) \sim \int d\sigma \exp (- {\sigma^2 \over {4 A}} + i Q \sigma),
\label{HS}
\end{eqnarray}
which introduces an auxiliary field $\sigma$.  A number of simplifications
now follow if one is interested only in those correlations among quark states
which contribute to chiral and diquark condensates.  We anticipate that the
saddle points of our non-linear sigma model will be described in the space of
auxiliary variables by dynamical masses that are real and independent of
chirality and flavor and by complex pairing potentials that are independent
of chirality.  In the following, we restrict our attention to such choices
and introduce accordingly real chiral and complex diquark fields $\sigma$
and $\Delta$.

Before proceeding, we also consider the possibility that a
spontaneous breaking of the color symmetry affects the chiral order
parameters.  This possibility clearly arises in the instanton model of color
superconductivity of Carter and Diakonov~\cite{instanton1}.  These authors
pointed out that the masses of the quarks with the condensing colors must be
different from those with the transverse colors if the Dyson-Gorkov equations
are to close.  To take this effect into account in the
present formulation, we include the projections of the original interaction
onto a chiral channel with color quantum numbers described by the diagonal
generator $\lambda_8$.  This introduces an additional Fierz constant $f_8^a$,
a term $f_8^a \,\delta_{ij'} (\lambda_8)_{\alpha\beta'} \times \delta_{i'j}
(\lambda_8)_{\alpha'\beta}$ on the right side of Eq.~(\ref{FierzRL}), and a
chiral field $\sigma_8$.  With these additions, the chiral fields couple to
quarks in the color diagonal combination $\psi^\dagger (\sigma+\sigma_8
\lambda_8) \psi$. This can also be written as $\psi^\dagger {\sigma_c} \psi$
where ${\sigma_c}\equiv {\rm diag}(\sigma_1,\sigma_1,\sigma_3)$ with
\begin{eqnarray}
\sigma_1\equiv \sigma+{\sigma_8 \over \sqrt{3}} \quad {\rm and} \quad  
\sigma_3 \equiv \sigma - 2 {\sigma_8 \over \sqrt{3}}.
\label{z8to13}
\end{eqnarray}
This final form of the fermion bilinear makes the separation in the chiral
fields explicit.  We follow this pattern in the external masses by taking
$m={\rm diag}(m_1,m_1,m_3)$.

The derivation of the sigma model potential is now straightforward.  Keeping
only the three auxiliary variables $\sigma_1$, $\sigma_3$, and $\Delta$, the
total four-fermion interaction ${\cal Y}$ becomes
\begin{eqnarray}
\exp{\cal Y} \sim \int d\sigma_1 d\sigma_3 d\Delta \,\, && \exp \left[ 
- 4 N \left( A |\Delta|^2 + B (2 \sigma_1+\sigma_3)^2/9 + 
C (\sigma_1-\sigma_3)^2/3 \right) 
\right. \nonumber \\
&& + \left.
\left(\begin{array}{c} \psi^\dagger_{1R} \\ \psi^T_{2R} \end{array}\right)
\left(\begin{array}{cc} 
  -\sigma_c & - \Delta \varepsilon \lambda_2 \\
  \Delta^* \varepsilon \lambda_2 & \sigma_c
      \end{array}
\right)
\left(\begin{array}{c} \psi^{\phantom\dagger}_{1R} \\ \psi^*_{2R} \end{array}
\right) + \left\{ {\rm R} \leftrightarrow {\rm L} \right\}
\right],
\label{expY}
\end{eqnarray}
where we have explicitly factored out chiral and spin factors in the terms
quadratic in auxiliary variables.  The coupling constants $A$, $B$, and $C$
are given as
\begin{eqnarray}
A \equiv 2 \left(\sum_{Ca} \,\, {f^a_\Delta \over
\Sigma^2_{Ca}}\right)^{-1}
\quad
B \equiv 2 \left(\sum_{Ca} \,\, {f^a_\chi \over 
\Sigma^2_{Ca}}\right)^{-1} 
\quad
C \equiv 2 \left(\sum_{Ca} \,\, {f^a_8 \over 
\Sigma^2_{Ca}}\right)^{-1}\ .
\label{ABdef}
\end{eqnarray}
Inserting then Eq.~(\ref{expY}) in
Eq.~(\ref{partfuncY}), we obtain the partition function of
Eq.~(\ref{partfunc}). Integrating out the fermion fields, we have 
\begin{eqnarray}
Z(T)&\sim& \int d\sigma_1\,d\sigma_3\,d\Delta \,\, \exp\left[-4 N
  \Omega(\sigma_1,\sigma_3,\Delta)\right], \label{ZOmega}
\end{eqnarray}
where the thermodynamic potential $\Omega$ is 
\begin{eqnarray}
\Omega(\sigma_1,\sigma_3,\Delta)&=&  
A |\Delta|^2 + B \left(\beta_1 \sigma_1^2 + \beta_2 \sigma_1 \sigma_3 +
\beta_3 \sigma_3^2\right) 
- (N_c-2) \log [(\sigma_3+m_3) ^2 + \pi^2 T^2] \nonumber \\
&& - 2 \log[(\sigma_1+m_1)^2+|\Delta+\eta|^2+\pi^2 T^2],
\label{Omega}
\end{eqnarray}
and where 
\begin{eqnarray}
\beta_1\equiv{4\over 9}+{1\over 3} C/B, \quad 
\beta_2\equiv{4\over 9}-{2\over 3} C/B, \quad 
\beta_3\equiv{1\over 9}+{1\over 3} C/B.
\label{betas}
\end{eqnarray}
Again, the external masses $m_1$ and $m_3$ and the parameter $\eta$ are to be
taken to zero at the end of the calculation.

The thermodynamic stable phases are given by the minima of the potential
$\Omega$ in Eq.~(\ref{Omega}). In practice, one does not expect any
condensation in repulsive channels; the global minimum must give zero
for the corresponding variables.  In the following we therefore consider the
$\Omega$ as given in Eq.~(\ref{Omega}) only in cases where the
chiral-$\lambda_8$ channel is attractive and $C$ is positive.  In cases where
the chiral channel $\lambda_8$ is repulsive, or $C<0$, we will explicitely
set $\sigma_1=\sigma_3$ as the field
$\sigma_8=(\sigma_1-\sigma_3)/\sqrt{3}$ is assumed not to condense.

In spite of the complications present in the initial formulation of this
model and its interactions, the potential $\Omega$ of Eq.~(\ref{Omega}) has a
very simple structure. In particular, we note that it depends only on three
coupling parameters, $A$, $B$, and $C$.  The arguments in the logarithms of
Eq.~(\ref{Omega}) represent squares of typical excitation energies in the
system.  Because of color mixing in the diquark channel, these excitations
lead in the case of two colors to a gap $\Delta$ in addition to the dynamical
mass $\sigma_1$.  The remaining colors develop a mass $\sigma_3$, which is in
general coupled to $\sigma_1$.  The question of whether chiral or color
symmetry breaking is preferred is now determined by the relative strengths of
the $\sigma$ and $\Delta$ fields.  This strength ratio depends solely on the
ratio $B/A$ or, equivalently, on the ratio of the Fierz projections of the
original interaction onto diquark and chiral channels, $B/A\sim f^{a}_\Delta/
f^{a}_\chi$. The ratio $C/B$ plays a secondary role, as it determines the
symmetry breaking pattern only when all three auxiliary fields acquire
non-zero mean field values.

In the large $N$ limit, the thermodynamically stable phases are determined by
the saddle points of the potential $\Omega$, Eq.~(\ref{Omega}).  We find that
$\Omega$ has precisely one local minimum for each fixed value of the coupling
parameters and the temperature.  There are thus only second order transitions
between single phases.  The variation of $B/A$, $C/B$, and $T$ reveals a rich
phase structure which is illustrated in Fig.~1.  Choosing the ratio $B/A$
fixes the slope of a straight line in parameter space passing through the
origin. Increasing the temperature $T$ corresponds to moving along that line
from the origin to large values of the axis coordinates.

It is useful at this stage to repeat our objective. The only interaction that
represents QCD is single-gluon exchange, which is a color octet, flavor
diagonal, and Lorentz vector interaction. This interaction realizes a ratio
$B/A=3/4$ for $N_c=3$. This correponds to the dot-dashed line in Fig.~1.  All
other interactions break at least one of the symmetries of QCD.  However, a
study of these deformations of QCD is useful on two respects. First, as
discussed shortly, it reveals that the phase diagram is mostly shaped by
color rather than chiral symmetry.  Second, as discussed in
Section~\ref{s:compare}, a comparison to the phase diagram of microscopic
models shows what aspects of the phase transition are protected by symmetry.
We now turn to a description of the phase diagram assuming that the
parameters $A$, $B$, and $C$, are chosen freely.  We will later consider what
ranges of values these parameters can actually take, given that they follow
from a single primary interaction, Eq.~(\ref{ABdef}).

We find the usual chiral phase transition whenever $B/A < N_c/2$. In
particular, $\Delta=0$ and $\sigma_1=\sigma_3$.  The potential $\Omega$ then
has the saddle-point solutions found in the $\chi$RMM at finite
temperature~\cite{JacksonVerbaar}.  The system breaks chiral symmetry
spontaneously at low temperatures $T < T_\chi$ with $T_\chi^2 = N_c/(\pi^2
B)$. Below $T_\chi$, we find chiral fields with a square root dependence in
the temperature, $\sigma_1=\sqrt{N_c/B}
\sqrt{1-T^2/T_\chi^2}$~\cite{JacksonVerbaar}.  Above $T_\chi$, chiral
symmetry is restored and $\sigma_1=\sigma_3=0$.  Spontaneous breaking of
color symmetry occurs only for ratios $B/A$ larger than $N_c/2$.  This region
of parameter space contains three phases.  First, large values of $B$ favor
chiral symmetric phases, $\sigma_1=\sigma_3=0$.  In this case, color is
spontaneoulsy broken below the critical temperature $T_\Delta=2/(\pi^2 A^2)$,
and is restored above. We again find a square root dependence for the pairing
gap below $T_\Delta$, $\Delta=\sqrt{2/A}\sqrt{1-T^2/T^2_\Delta}$. Second,
keeping $A$ fixed to values $A< 2/(\pi^2 T^2)$ and decreasing $B$ from large
values, one encounters a phase of mixed symmetry breaking where all
condensates are developed and where $\sigma_1 \neq \sigma_3$. The upper
boundary of this phase depends on $C$; we have plotted three examples of this
critical line in Fig.~1.

Before further discussing the nature of the mixed symmetry breaking phase, we
must distinguish cases where the chiral-$\lambda_8$ channel is attractive and
$C>0$ from those where it is repulsive and $C<0$. In the repulsive cases, we
argued before that $\sigma_1=\sigma_3$. The upper boundary is then the line
$B = A + (N_c-2)/ \pi^2 T^2$. Below that line and above the line $B/A =
N_c/2$, we again find square root dependence for both chiral and diquark
fields.  Attractive cases ($C>0$) require more care.  Anticipating the
following longer discussion of those regions of parameter space which are
accessible, we remark that there are strict limits to the parameter ratio
$C/B$ which can be realized.  We choose $N_c=3$ for the present discussion.
A color octet interaction gives a ratio $C/B=f^O_\chi/f^O_8=-16/3$; a color
singlet leads to $C/B=f^S_\chi/f^S_8=2/3$.  It is therefore clear from the
definition of $B$ and $C$, Eq.~(\ref{ABdef}), that any combination of octet
and singlet interactions which produces a positive ratio $C/B$ must also
satisfy $C/B \ge 2/3$.

The pattern of symmetry breaking now depends on the interaction. Let us start
by the case $C/B=2/3$, realized by a color singlet interaction.  The minimum
of $\Omega$ occurs for $\sigma_1=0$ and, since $\beta_2=0$
(Eq.~(\ref{betas})), $\sigma_3$ decouples from $\Delta$. The diquark field
exhausts the strength alloted to the condensing colors, and chiral symmetry
breaking acts independently on the third color.  The upper boundary for this
symmetry breaking pattern is the horizontal line $B=3/(\pi^2 T^2)$.  Let us
now increase $C/B$ above $2/3$.  The upper boundary moves down as illustrated
in Fig. 1 and reaches the line $B=A+(N_c-2)/(\pi^2 T^2)$ in the limit $C/B
\to \infty$.  For any intermediate value of $C/B \le 2/3$, the chiral fields
inside the wedge of mixed symmetry breaking satisfy $\sigma_3/\sigma_1=1+(3
-2 A/B)/(C/B-2/3)$.  The sigma fields have again a square root dependence on
the temperature.

It is remarkable that the overall structure of the phase diagram depends
primarily on the critical ratio $B/A \sim N_c/2$. This follows because chiral
condensation involves all $N_c$ colors, while quark Cooper pairs mix two
colors only.

Now comes the question of understanding whether one can really access the
richness of the phase diagram of Fig. 1 with a random matrix model. We noted
above that there are strict limits on the parameter ratios which can actually
be realized.  As we shall demonstrate in the following section, the upper
limit on the ratio $B/A$ is actually $N_c/2$.  This precludes the exploration
of phases with $\Delta \neq 0$, and thus none of the random matrix model can
support stable diquark condensates at zero chemical.  It is possible to come
close to a stable diquark phase in the case of an axial or a tensor
interaction in the color singlet channel.  In this limiting case, the diquark
condensate can always be rotated into the chiral condensate for colors $1$
and $2$, while the remaining $N_c-2$ colors break chiral symmetry
independently.  Even in this extreme case, the diquark condensate does not
represent an independent thermodynamic phase.

\section{Allowed and forbidden regions of parameter space}
\label{s:results}

We now wish to demonstrate that the maximum attainable ratio $B/A$ is
$N_c/2$.  The quantity $B/A$ is a weighted ratio of the Fierz coefficients of
the original interaction in chiral and diquark channels, Eq.~(\ref{ABdef}).
These coefficients were obtained from projecting certain chiral blocks, see
Eqs.~(\ref{FierzRL}) and (\ref{FierzRLT}). Let us first consider a given
Lorentz channel $C$, described by $n_C$ matrices $\Gamma_C$ ($n_C=1,1,4,4,6$
when $C=S,P,V,A,T$, respectively). Its Fierz coefficient in the chiral
channel is $f^{a}_\chi = (n_C/2) c^a$, where the color factor $c^a$ is
$c^S=1/N_c$ for a color singlet and $c^O=2 (N^2_c-1)/N_c^2$ for a color
vector interaction, respectively.  The Fierz coefficent in the diquark
channel is
\begin{eqnarray}
f^{a}_\Delta &=& - 2\,\, \sum_{\mu=1,n_C}
\frac{{\rm Tr}[(C \gamma^5) \Gamma_C^{\phantom T} (C \gamma^5) \Gamma_C^T]}
{{\rm Tr}[(C \gamma^5)^2]^2}\,\,
\frac{{\rm Tr}[(i \lambda_2) \Lambda_a^{\phantom T} (i \lambda_2) \Lambda_a^T]}
{{\rm Tr}[\lambda^2_2]^2},
\label{fLorentz}
\end{eqnarray}
where $\sum_{\mu=1,n_C}$ is a sum over the $n_C$ members of the Lorentz
channel. The ratio of traces of Dirac matrices can be determined from their
transposition property. Noting that
\begin{eqnarray}
(C \gamma^5)\, \Gamma_C^T\, (C\gamma^5) &=& - \Gamma_C^{\phantom T} \quad
{\rm when} \,\,C=S,P,V, \\
(C \gamma^5)\, \Gamma_C^T\, (C\gamma^5) &=& + \Gamma_C^{\phantom T} \quad
{\rm when} \,\,C=A,T,
\end{eqnarray}
we deduce that ratio to be
$-n_C/4$ when $C=S,P,V$ and $n_C/4$ when $C=A,T$. The ratio of traces of
color matrices can be deduced 
from the completeness relations for the generators of $SU(N)$,
\begin{eqnarray}
\sum_{a=1}^{N_c^2-1} \lambda^a_{\alpha\beta}\,\,\lambda^a_{\gamma\delta}&=&
2 \left(\delta_{\alpha\delta}\,\,\delta_{\beta\gamma}-{1\over N_c}
\delta_{\alpha\beta}\,\,\delta_{\gamma\delta}\right).
\end{eqnarray}
Thus, color contributes a factor $f^{C|S}_\Delta \sim -1/2$ for a
singlet and $f_\Delta^{C|O} \sim 1+1/N_c$ for a vector interaction,
respectively. For reference, we note that single-gluon exchange produces a
ratio of Fierz coefficients $f^O_\Delta/f^O_\chi=3/4$.

We now combine several Lorentz and color channels and try to obtain the
largest ratio $B/A$. Large ratios are obviously realized by attractive
quark-quark interactions. These channels have positive coefficients
$f_\Delta^a$. They correspond to a Lorentz scalar, pseudoscalar, or vector
with a color interaction in the $N_c^2-1$ channel or to a Lorentz tensor or
pseudovector with a singlet color interaction.  Their combination gives a
ratio
\begin{eqnarray}
{B \over A} &=& \frac{
\left(\Sigma_{S|O}^{-2}+\Sigma_{P|O}^{-2}+4 \Sigma^{-2}_{V|O}\right)   
{(N_c+ 1)/N_c}
+ \left(4 {\Sigma_{A|S}^{-2}} +6 \Sigma_{T|S}^{-2} \right){1/2}
}{
\left({\Sigma_{S|O}^{-2}}+\Sigma_{P|O}^{-2}+4 \Sigma_{V|O}^{-2}\right) 
{2 (N_c^2-1)/N_c^2}
+ \left(4 {\Sigma_{A|S}^{-2}}+6 \Sigma_{T|S}^{-2} \right) {1/N_c}
} 
\label{main}
\end{eqnarray}
It is now clear that $B/A$ is bounded from above as $B/A \le N_c/2$, for $N_c
\ge 2$. The upper bound can be reached by keeping only the last terms in
numerator and denominator.  This corresponds to either an axial or tensor
interaction in a color singlet channel.

It is instructive to consider the phase diagram of a model with the limiting
value $B/A=N_c/2$. We noted earlier that the coupling between chiral fields
vanishes when the interaction is a color singlet: now, $C/B=2/3$ which
implies $\beta_2=0$, see Eq.~(\ref{betas}). The phase diagram follows
immediately.  There is only one critical temperature $T_c= 2 /(\pi^2 A)$
below which chiral and color symmetries are broken, and their fields satisfy
$\sigma_3^2=\Delta^2+\sigma_1^2=(2/A)(1-T^2/T_c^2)$.  The symmetry breaking
pattern for color is thus $SU(N_c)\to SU(2)\times SU(N_c-2)$.  Colors $1$ and
$2$ exhibit a diquark condensate, while chiral symmetry operates
independently on the third color. However, one can rotate a given color
broken solution $(\Delta,\sigma_1)=(\Delta_0,0)$ into
$(\Delta,\sigma_1)=(0,\Delta_0)$.  This rotation brings the original state to
one for which $\sigma_1=\sigma_3$ and $\Delta=0$. The initial state is thus
equivalent to one of broken chiral symmetry, and the diquark condensate does
not describe a thermodynamically independent phase.  We show in the appendix
that the chiral and diquark order parameters can be described by Banks-Casher
formulas. However, because of the rotational symmetry, all three order
parameters are actually related to the same spectral correlator.

We briefly return to the issue of implementing more flavor structure in the
Dirac operator by including the flavor generators $\tau_i$.  The additional
flavor symmetry would contribute to the Fierz projections in the chiral and
diquark channels. However, these additional factors would be those of an
$SU(2)$ symmetry, and would therefore be equal in both channels.  An
additional flavor structure in the interaction cannot raise the ratio $B/A$
above $N_c/2$.

\section{Comparison with a microscopic model}
\label{s:compare}

The phase diagram that we have obtained shares many features with those of the
microscopic model of Berges and Rajagopal~\cite{BR}, and other NJL
models~\cite{NJLsuper}. We can summarize the three main ingredients of these
models as follows.  First, the four-fermion interaction respects the global
chiral and color symmetries of QCD.  Second, the coupling constants in the
chiral and diquark channels are left as free parameters; particular choices
of these parameters correspond to an interaction induced either by instantons
or by single-gluon exchange.  Third, the effects of asymptotic freedom are
implemented by form factors whose actual form is model-dependent.

A comparison bewteen NJL-models and the present matrix models clearly shows
what predictions on the phase diagram are protected by symmetry. To
illustrate this, let us start from the effective potential given in
Ref.~\cite{BR}, and take the limit of zero chemical potential. With a few
obvious renaming of variables, we obtain
\begin{eqnarray}
\Omega&=&A \Delta^2+B \sigma^2 -2 \int \frac{q^2 dq}{\pi^2} \bigg\{
(N_c-2) \Big\{E_\sigma+2 T \log \Big(1+\exp[-E_\sigma/T] \Big)
\Big\}
\nonumber \\
&& +2\Big\{E_\Delta+2 T \log \Big(1+\exp[-E_\Delta/T]
\Big)\Big\}\bigg\},
\label{omegaBR}
\end{eqnarray}
where the single particle energies are $E_\sigma^2\equiv q^2+F^4 \sigma^2$,
$E_\Delta^2\equiv q^2+ F^4 (\sigma^2+\Delta^2)$, and $F
=\Lambda_{\rm qcd}^2/(q^2+\Lambda_{\rm qcd}^2)$ is an {\it adhoc} form factor.
Following Berges and Rajagopal, we do not consider here that the chiral
condensates can split in color space, and we thus retain a single chiral field,
$\sigma_1=\sigma_3=\sigma$.  Remarkably, the potential $\Omega$ gives a phase
diagram identical to that of Fig.~1.  We find again a single minimum to
$\Omega$ for each set of parameters. This results in single phases separated
by straight line boundaries of second order transitions. The general topology
of the phase diagram is mostly set by color symmetry.  The overall scale is
however sensitive to form factors.  Let us for instance choose the coupling
constants such that $B/A<N_c/2$.  For these ratios, our model has one
critical line $B = N_c/(\pi^2 T^2)$ where chiral symmetry is restored.  This
line translates in the NJL model into the line $B=N_c {\cal F}(T)$,
where
\begin{eqnarray}
{\cal F}(T)\equiv 2 \int\frac{dq q}{\pi^2}
F^4 \tanh[{q\over 2 T}]
\end{eqnarray}
Thus the scale $1/(\pi^2 T^2)$ translates into ${\cal F}(T)$, which
is sensitive to the particular choice of the form factor $F$. This of course
implies that the actual value of the critical temperature is model-dependent.
But the topology of the phase diagram, and in particular the line $B=N_c
A/2$, is protected by symmetry.

Our approach can also be compared to the O(N) symmetric
two-dimensional model of Chodos et al.~\cite{Chodos}.  This model is a
generalization of the Gross-Neveu interaction to include pairing forces.
Repeating the discussion of the previous section, one would infer a phase
diagram similar to that of Fig.~1. Given the interaction of
Ref.~\cite{Chodos}, the critical line $B=A N_c/2$ become
$A-B=0$. In Ref.~\cite{Chodos}, $A-B$ actually corresponds to a parameter
$\delta$ which determines which of the chiral ($A-B>0$) or the diquark
($A-B<0$) condensates exists in the vacuum.  The parameter $\delta$ is
further shown to be invariant under renormalization. This result is
consistent with the fact that the topology of the phase diagram only depends
on global symmetries.

Although the present discussion indicates that the random matrix and NJL-type
microscopic approaches share fundamental features, they differ in one
essential point. NJL-studies generally assume from the start a potential of
the form of Eq.~(\ref{Omega}) on the basis of symmetry considerations. The
coupling contants $A$, $B$, and $C$, may be related to one another in
specifically motivated cases, but are in general taken as free parameters
whose range of variation is dictated by the phenomenology to be reproduced.
By contrast, random matrix models start out at a more microscopic level.  We
first selected certain types of interactions satisfying global symmetries,
and then mimicked the dynamics by means of a random background. We also
confined ourselves to Hermitean matrix models, in which a meaningful relation
between order parameters and spectral properties can be found. As a result of
working at a microscopic level, the coupling parameters of the potential in
Eq.~(\ref{Omega}) satisfy strong constraints such as $B/A< N_c/2$.  These
constraints are not deduced from phenomenology but are now inherited from the
dynamics of the primary interactions which have been integrated over.
Starting from a more microscopic level thus has the clear advantage of
implementing those dynamical constraints, while still permitting the
construction of the desired global symmetries.

When comparing with microscopic models, it is also necessary to consider
exchange effects. Random matrix models include only direct terms. Exchange
terms involve quark fields with unequal matrix indices $k$ and $l$ in
Eqs.~(\ref{vacurrent}) and (\ref{sptcurrent}), and are therefore $1/N$
suppressed. To take exchange effects explicitely into account, we must
modify by hand the four-fermion interaction produced by integration over the
random background. This amounts to adding to the original four-fermion
interaction its Fierzed transformed expression~\cite{NJLreview}. One way of
proceeding is to fine tune the variances of the random background in order to
reproduce the modified four-fermion interaction. It turns out that only
scalar interactions produce exchange effects that alter the chiral channel;
exchange actually changes the sign of the Fierz coefficient in that channel.
The modified interaction is repulsive for chiral condensation and may thus
favor diquark condensation.  This opens the possibility of exploring
the region of the phase diagram above the line $B/A > N_c/2$.  However, the
matrix model that would mimic exchange is non-Hermitean, a case which is
oustide the scope of this paper.

A final comment concerns the similarities between single-gluon exchange and
instanton-induced interactions in regard with the explored region of the
phase diagram. To mimic single-gluon exchange we choose a vector interaction,
which produces $B/A=3/4$.  To understand how to mimic an instanton-induced
interaction, we first notice that it conserves chirality (we assume here an
interaction arising by scattering on single instantons). Thus, a genuine
matrix model would combine scalar, pseudoscalar, and tensor interactions, and
would then be very different in character from single-gluon exchange. It
turns out that an instanton-induced interaction actually produces the same
ratio $B/A=3/4$ as single gluon exchange, provided exchange effects are taken
into account~\cite{instanton2}. This result is not surprising. The
instanton-induced interaction is generated by integration over a classical
gluon background which couples to quarks as a color vector. A consistent
treatment must remember the vector nature of the interaction and must produce
a ratio $B/A=3/4$. Therefore, QCD and any of its consistent approximations
should lie on the line $B/A=3/4$.
 
\section{Conclusions}

In this paper, we have introduced a random matrix model that can in principle
admit both chiral and diquark condensation and have studied the competition
of these two forms of order at zero chemical potential. We have considered
interactions that are Lorentz invariant in the vacuum and gauge symmetric
and have displayed the rich structure of the ensuing phase diagram in
parameter space. The topology of the phase diagram is mostly governed by
color symmetry. Its exploration requires the variation of only two ratios of
Fierz coupling constants. We have further argued that there exist strong
constraints on the values that these ratios can actually achieve. For
interactions represented by Hermitean matrices, the ratio of the coupling
constant in the chiral channel to that in the diquark channel is necessarily
less than the critical value $N_c/2$ required to favor the formation of
Cooper pairs. Thus, none of the present random matrix models can support
stable diquark condensates.

Our arguments are primarily based on symmetry considerations that also apply
to QCD. This leads us to the conclusion that no mean-field treatment of QCD
with two light flavors can support independent diquark condensates at zero
chemical potential.  This conclusion does of course respect the
phenomenological evidence that the QCD vacuum spontaneously breaks chiral
symmetry, but not color symmetry. The main message of our approach, however,
is to emphasize that the coupling constants of an effective potential are
constrained by the underlying symmetries of the interaction they mean to
represent. Determining these constraints in a given physical situation 
helps rule out certain forms of order on the basis of symmetry alone.

\section*{Appendix} 
\label{s:axial}

In this appendix, we explore further the case of an axial interaction with a
color singlet structure. As we showed in Section~\ref{s:results}, this
interaction achieves a ratio $B/A=3/2$. We found before that the gap
equations yielded solutions with a finite diquark condensate which are
degenerate with those of pure broken chiral symmetry.  We now wish to derive
Banks-Casher relations for the condensates.  We show that the degeneracy of
the solutions forces all condensates to be related to the same component of
the Dirac spectrum.

The two chiral and the diquark condensates can be obtained from the partition
function by taking 
\begin{eqnarray}
\langle \psi^\dagger_{1 \alpha} \psi_{1 \alpha}^{\phantom\dagger}
\rangle &=& \lim_{m_1 \to 0} \lim_{N\to \infty} {1\over 2 N N_f} 
\left.{\partial \log Z(T) \over \partial m_1} \right|_{\eta=0,m_3=0}
{\rm when} \,\,\alpha=1,2, \label{dlogone} \\
\langle \psi^\dagger_{1 \alpha} \psi_{1 \alpha}^{\phantom\dagger}
\rangle &=& \lim_{m_3 \to 0} \lim_{N\to \infty} {1\over  N N_f} 
\left.{\partial \log Z(T) \over \partial m_3} \right|_{\eta=0,m_1=0}
{\rm when } \,\,\alpha=3, \label{dlogtwo} \\
\langle \psi^T_{2} P_\Delta \psi_{1}^{\phantom\dagger}
\rangle &=& i \lim_{\eta \to 0} \lim_{N\to \infty} {1\over  N} 
\left.{\partial \log Z(T) \over \partial \eta^*} \right|_{m_1=0,m_3=0}. 
\label{dlogthree}
\end{eqnarray}
Here, as usual, it is important to take the thermodynamic limit $N\to \infty$
first to obtain physically meaningful quantities for the
condensates~\cite{JacksonVerbaar}.
The derivatives can be evaluated in a formal way. Integrating the
partition function in Eq.~(\ref{partfunc}) over the random background, an
operation which we represent here by brackets, we  obtain
\begin{eqnarray}
Z(T)&=&\left\langle \exp {\rm Tr} \log i {\cal D} \right\rangle
\end{eqnarray}
where ${\cal D}$ is the flavor block matrix
\begin{eqnarray}
{\cal D} &=& \left(
\begin{array}{cc}
{\cal H}+{\cal T}+i m & \eta P_\Delta \\
-\eta^* P^\dagger_\Delta & -{\cal H}^T+{\cal T}-im \\
\end{array}
\right).
\end{eqnarray}
Taking the derivatives in Eqs.~(\ref{dlogone}), (\ref{dlogtwo}), and
(\ref{dlogthree}), we have
\begin{eqnarray}
\langle \psi^\dagger_{1 \alpha} \psi_{1 \alpha}^{\phantom\dagger}
\rangle&=& \lim_{m_1 \to 0} \lim_{N \to \infty} {i\over N N_f} 
\left.
\left\langle \frac{{\rm det} i {\cal D}}{\langle {\rm det} i {\cal D}\rangle}
{\rm Tr} \left[P_1
\left( {\cal D}^{-1}_{11}-{\cal D}^{-1}_{22}\right)\right]
\right \rangle \right|_{\eta=0,m_3=0} \quad (\alpha=1,2),\\
\langle \psi^\dagger_{1 \alpha} \psi_{1 \alpha}^{\phantom\dagger}
\rangle&=& \lim_{m_3 \to 0} \lim_{N \to \infty} {i\over N N_f}
\left. 
\left\langle \frac{{\rm det} i {\cal D}}{\langle {\rm det} i {\cal D}\rangle}
{\rm Tr} \left[P_3 
\left( {\cal D}^{-1}_{11}-{\cal D}^{-1}_{22}\right)\right]
\right \rangle
\right|_{\eta=0,m_1=0}\quad (\alpha=3),\\
\langle \psi^T_2 P_\Delta \psi_1^{\phantom\dagger} 
\rangle&=& \lim_{\eta \to 0} \lim_{N \to \infty} {-i\over N} \left. 
\left\langle \frac{{\rm det} i {\cal D}}{\langle {\rm det} i {\cal D}\rangle}
{\rm Tr} \left[P_\Delta {\cal D}^{-1}_{12}\right] 
\right \rangle \right|_{m1=0,m3=0},
\end{eqnarray}
where the color operators $P_1={\rm diag}(1,0,0) $ and $P_3={\rm
  diag}(0,0,1)$ project onto colors $1$ and $3$ respectively.  To evaluate
the matrices ${\cal D}^{-1}_{11}$, ${\cal D}^{-1}_{22}$, and ${\cal
  D}^{-1}_{12}$, one makes use of the fact that ${\cal H}$ is color diagonal,
${\cal H}={\rm diag}(H,H,H)$, and that $H$ satisfies $C \gamma^5 \,H^T \,C
\gamma^5=H$. A diagonalization of the flavor block matrices gives then
\begin{eqnarray}
\langle \psi^\dagger_{1 \alpha} \psi_{1 \alpha}^{\phantom\dagger}
\rangle &=& \lim_{m_1 \to 0} \lim_{N \to \infty} {1\over N} 
\left\langle \frac{{\rm det} i {\cal D}}{\langle {\rm det} i {\cal D}\rangle}\,
\frac{m_1}{({H}+{\cal T})^2+m_1^2} \right\rangle \quad (\alpha=1,2),\\
\langle \psi^\dagger_{1 \alpha} \psi_{1 \alpha}^{\phantom\dagger}
\rangle &=& \lim_{m_3 \to 0} \lim_{N \to \infty} {1\over N} 
\left\langle \frac{{\rm det} i {\cal D}}{\langle {\rm det} i {\cal D}\rangle}\,
\frac{m_3}{({H}+{\cal T})^2+m_3^2} \right\rangle \quad (\alpha=3), \\
\langle \psi^T_2 P_\Delta\psi_{1}^{\phantom\dagger}
\rangle &=& i \lim_{\eta \to 0} \lim_{N \to \infty} {1\over N} 
\left\langle \frac{{\rm det} i {\cal D}}{\langle {\rm det} i {\cal D}\rangle}\,
\frac{\eta}{({H}+{\cal T})^2+|\eta|^2} \right\rangle,
\end{eqnarray}
We thus see that the three order parameters are related to the same spectral
properties of the Dirac operator. This result confirms the degeneracies of
the solutions with finite diquark fields with those of broken chiral
symmetry.

\section*{Acknowledgements}
We thank G. Baym, G. Carter, K. Splittorff and J. J. M. Verbaarschot for very
useful discussions.


\newpage

\begin{figure}[b]
\setlength\epsfxsize{15cm} \centerline{\epsfbox{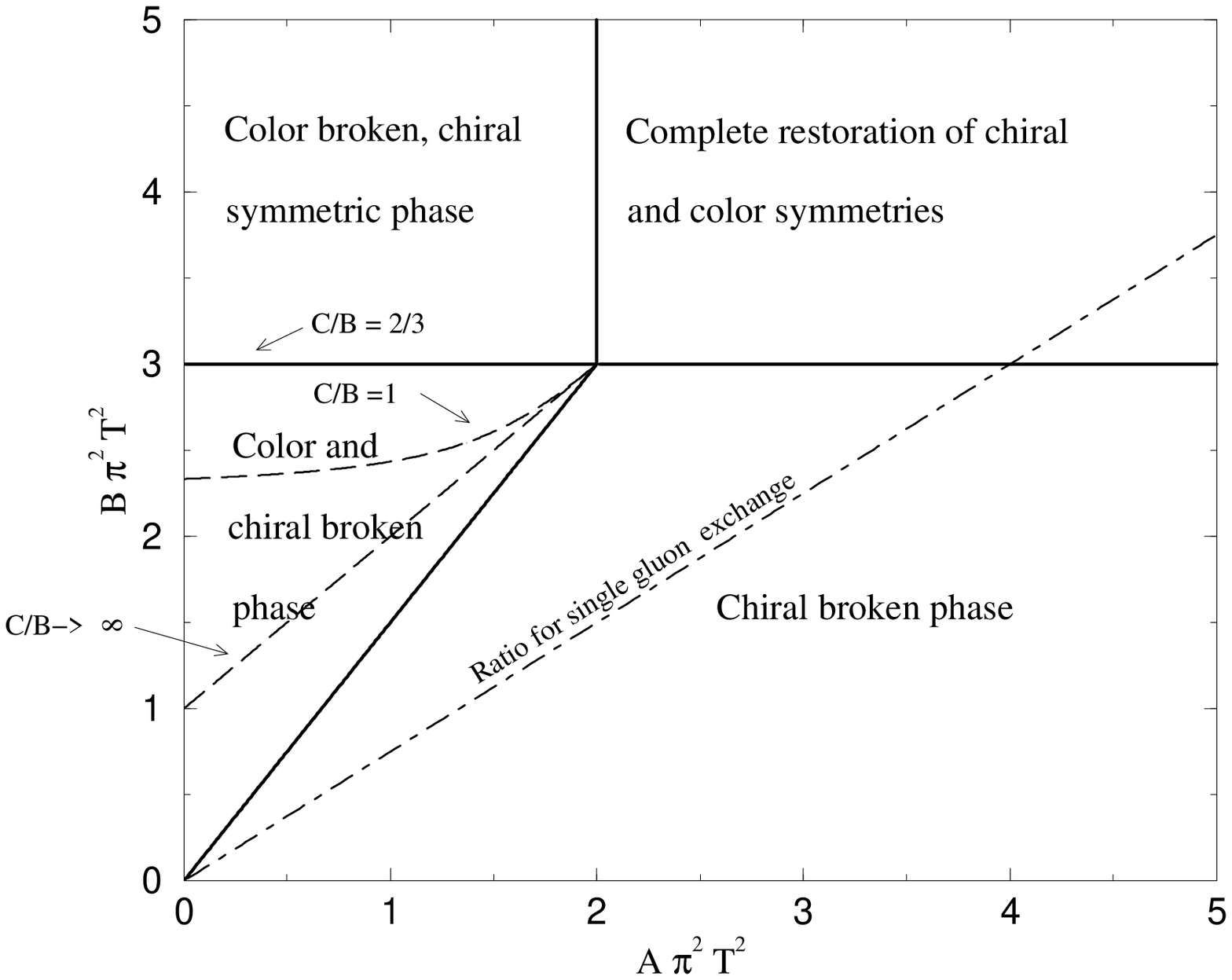}}
\caption{
  The phase diagram in parameter space. The horizontal axis represents the
  coupling constant in the diquark channel, the vertical axis represents that
  in the chiral channel. The solid curves are second-order phase transitions.
  The long-dashed lines correspond to the upper boundary of the phase where
  both color and chiral symmetries are broken. We show this boundary for
  various values of the ratio of coupling constants in the chiral and the
  chiral-$\lambda_8$ channels (see text). The dot-dashed line is the line of
  constant ratio $B/A=3/4$ obtained with single gluon exchange and $N_c=3$.
  This line does not cross any phase with a finite diquark condensate.}
\end{figure}

\end{document}